# CanvOI, an Oncology Intelligence Foundation Model: Scaling FLOPS Differently

Jonathan Zalach[1,*], Inbal Gazy[1,*], Assaf Avinoam[1], Ron Sinai[1], Eran Shmuel[1], Inbar Gilboa[1], Christine Swisher[2], Naim Matasci[3], Reva Basho[3] and David B. Agus[3]

[1] Imagene   [2] Oracle   [3] Ellison Institute of Technology

## Abstract

The rapidly evolving field of digital oncopathology faces significant challenges, including the need to address diverse and complex clinical questions, often involving rare conditions, with limited availability of labeled data. These limitations hinder the development of robust AI-driven tools in the biomedical space, where accuracy in probabilistic determinations is of utmost importance. To address this, digital pathology foundation models have begun to emerge, typically developed with the size and diversity of the pre-training dataset and model parameters in mind. Here, we present CanvOI, a ViT-g/10-based foundation model designed to enhance the capabilities of digital pathology by addressing these challenges through a different approach. Considering the unique nature of oncologic histopathological images and the requirements from the embeddings to provide meaningful representations for Multiple Instance Learning (MIL) downstream models, we chose to modify the input image characteristics. By introducing larger tile sizes ($380^2$ pixels) and smaller patch sizes ($10^2$ pixels), we were able to optimize the model's performance, pushing computational resources in a new direction and achieving state-of-the-art performance on cancer-related benchmarks. CanvOI demonstrated a 1.5-7.4% improvement in averaged AUC compared to other leading foundation models built for digital pathology. Moreover, our results demonstrate that CanvOI significantly outperformed the other models, with the performance gap widening substantially when trained on just 10% of the initial cohort. This work highlights an alternative approach that, if integrated with traditional development approaches, has the potential to advance Oncology Intelligence (OI), overcome some of the current barriers and ultimately improve the clinical outcome of cancer patients.

## Introduction

Histopathology is fundamental to clinical medicine and biomedical research, providing essential information for understanding disease pathogenesis and making informed clinical decisions. With the advancements in artificial intelligence (AI) applications in digital pathology, the vast amount of data encapsulated within pathology images is continuously being uncovered (for example[1-6]) indicating the enormous potential yet to be realized. While AI has made significant inroads in digital pathology, challenges such as generalizing across diverse datasets and addressing rare conditions with limited data remain prominent[7,8]. Moreover, the availability and access to high quality images with labeled data is restricted in the biomedical domain, limiting the ability to develop accurate AI models for specific applications.

---

* Equal contribution

 Corresponding author: jonathan@imagene-ai.com

To overcome these hurdles, computer vision foundation models are emerging at the forefront of AI, holding the promise to become leading solutions[9-11]. Foundation models, pre-trained on large unlabeled histological datasets, generate embeddings, vector representations of the complex features and patterns within the images. Unlike task-specific models, foundation models are versatile and can be adapted to a wide range of downstream applications with significantly less labeled data than typically required. By reducing dependency on extensive data curation, these models can enable the exploration of previously unattainable research areas and open up transformative possibilities for medical research and diagnostics, particularly with image-based data. Although they hold great potential, foundation models are still in the early stages in the realm of digital pathology. Significant efforts are underway to develop foundation models that address biomedical needs, aiming to achieve robust and accurate outcomes across a broad range of diagnostic, prognostic, and theragnostic questions, enabling their wide utilization in various medical contexts[12-14].

When designing foundation models for the medical field, three critical and interdependent scaling factors should be considered: the diversity and size of the pre-training dataset, model size and computational power. First, the amount of data and diversity the model is exposed to during pre-training is critical in creating a strong foundation model that captures the broad and various patterns and complexities of pathological images. This is of paramount importance, since, unlike in some disciplines, the biomedical space has little tolerance for instability due to unforeseen artifacts or image variabilities. Furthermore, a more extensive pre-training image dataset, with a wider variety of sample types, processing methods and acquisition conditions enhances the model's ability to support a range of downstream tasks, including those that address disease heterogeneity and tackle less common traits. To develop models that accurately reflect clinical realities, it is vital to include data that covers a broad spectrum of images with variations found in the real-world, such as different artifacts, digital slide scanners, sample collection, sample processing and staining techniques. This comprehensive approach ensures that models would be able to navigate the inherent complexities and messiness of real-world medical data, while maintaining stability and accuracy. Initially, The Cancer Genome Atlas (TCGA) served as the primary dataset for pre-training foundation models[15,16]. Over time, the quantity and diversity of data used for training these models have expanded, leading to better-performing models and improved adaptability across various medical applications[14,17]. Second, the size of the model significantly influences its stability. Research has demonstrated that larger models, which can encapsulate more detailed and intricate patterns, yield more reliable and stable outcomes showing enhanced performance[9,18] which is crucial in the biomedical field. These models are better equipped to generalize across diverse datasets and address rare conditions with limited data availability, thereby pushing the boundaries of what is possible in medical diagnostics and research. Zhai et al.[19] showed that smaller models have limited capacity, preventing them from fully utilizing larger datasets and compute resources. In contrast, larger models are more sample-efficient and can leverage extensive datasets. The capacity of large models to process and learn from more extensive data contributes to their robustness and reliability[19,20]. Finally, when supporting a sufficiently large pre-training dataset, an increase in computational power can lead to better performing models. However, the significant compute requirements for training large models can be a barrier, making it crucial to select the appropriate model based on the available data, hardware capabilities and compute budget[21].

These scaling factors indeed play a crucial role in the ability of foundation models to produce more powerful and meaningful embeddings, often receiving significant attention in model optimization. While such factors pose significant considerations, other factors should also be taken into account to enhance these models for histological image-related tasks. One such factor is the vision transformer patch size. Studies have shown that decreasing the patch size generally enhances performance[9,22,23]. However, similarly to increasing the number of model parameters, reducing the patch size also comes with higher computational costs. Yet, while increasing the size of Vision Transformer (ViT) models beyond certain thresholds can lead to instability during training, often resulting in neural networks that are challenging to train[18], our experiments suggest that reducing the patch size can be done without significantly destabilizing the model, at least up to a certain scale. While we did observe slight instability when reducing to smaller patch sizes and increasing the input image size, these issues were manageable.

Taking all this under consideration, we developed CanvOI, a Cancer Vision Oncology Intelligence foundation model. CanvOI is a 1.1B parameter model that processes, unlike the commonly used images, tiles the size of $380^2$ pixels divided into patches of $10^2$ pixels (ViT-g/10). This approach allowed us to push the computational demands in digital pathology to new heights and achieve state-of-the-art performance on cancer-related benchmarks, suggesting that this method could guide future development of digital pathology foundation models to enhance their capabilities.

# Methods

## CanvOI Architecture and Training

The pre-training pathology dataset consisted of 632,608 tissue samples with 70,217,688 tiles, extracted from the tissue-containing regions of the images. The dataset included almost exclusively hematoxylin and eosin (H&E)-stained tissue images, from over 100 international source sites and covering more than 40 major organs and tissue types (Figure 1). The vast majority (more than 90%) of our extensive and diverse dataset included samples from our internal dataset and the rest from publicly available datasets including TCGA[24] (data available at: https://portal.gdc.cancer.gov/), CPTAC[25] (data available at: https://www.cancerimagingarchive.net/), GTEx[26] (data available at: https://www.gtexportal.org/home/), CMB[27] (data available at: https://www.cancerimagingarchive.net/) and CAMELYON17[28] (data available at: https://grand-challenge.org/). The images were obtained using dedicated digital pathology scanners, scanned at either 20× or 40× resolution. All images were processed to 20× magnification (0.5 μm/pixel).

Our model, CanvOI, is built on a ViT-g/10 architecture and utilizes the DINOv2 framework[9]. For pre-training, tissue images were extracted at a size of 670 × 670 pixels. From these, smaller crops were extracted using ranges of [0.2-0.57] and [0.03-0.2] for global and local views respectively and resized to 380 × 380 pixels for global views and 140 × 140 pixels for local views (Figure 2).

To optimize efficiency, we initially trained the model with 224 × 224 pixel tiles and later fine-tuned it using 380 × 380 pixel tiles. Previous studies have shown that the performance drop with this approach is minimal compared to training with 380 × 380 pixel tiles from the start[29].

To reduce noise, a set of non-overlapping tissue-containing tiles was created for each slide, filtering out tiles that did not meet a predefined threshold for tissue presence. This filtering was done using Canny edge detection[30] and color averaging techniques. The resulting filtered sets were then used as input for the AB-MIL model's training and evaluation.

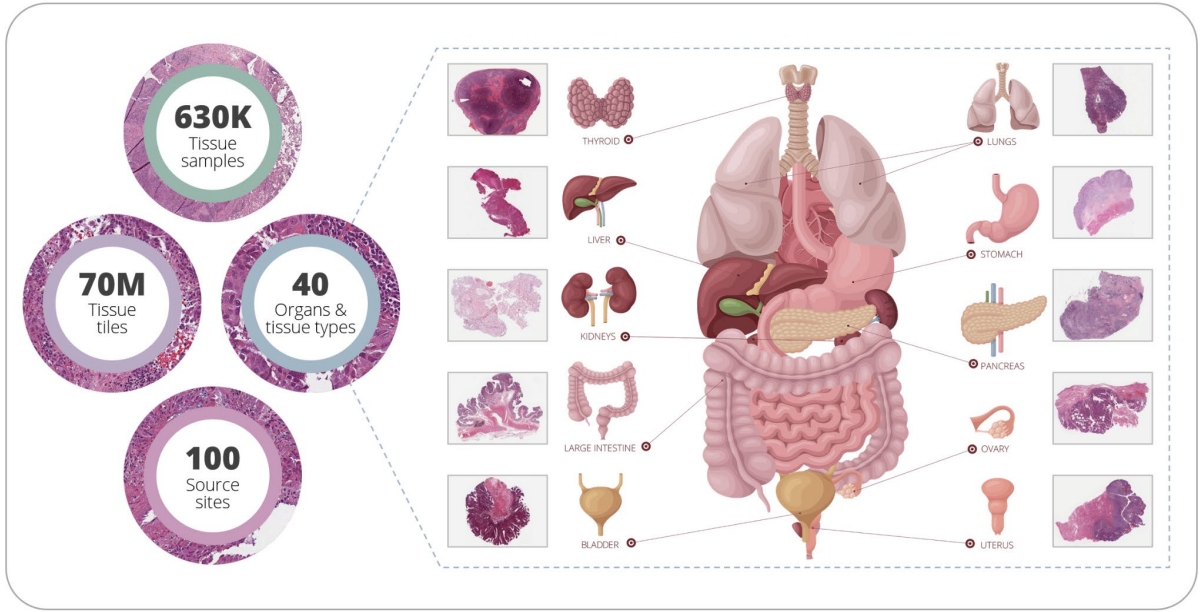

**Figure 1. CanvOI Pre-training dataset characteristics.**
**Left:** Description of the dataset according to the different features. **Right:** Visual representation of WSIs from selected organs included in the pre-training dataset.

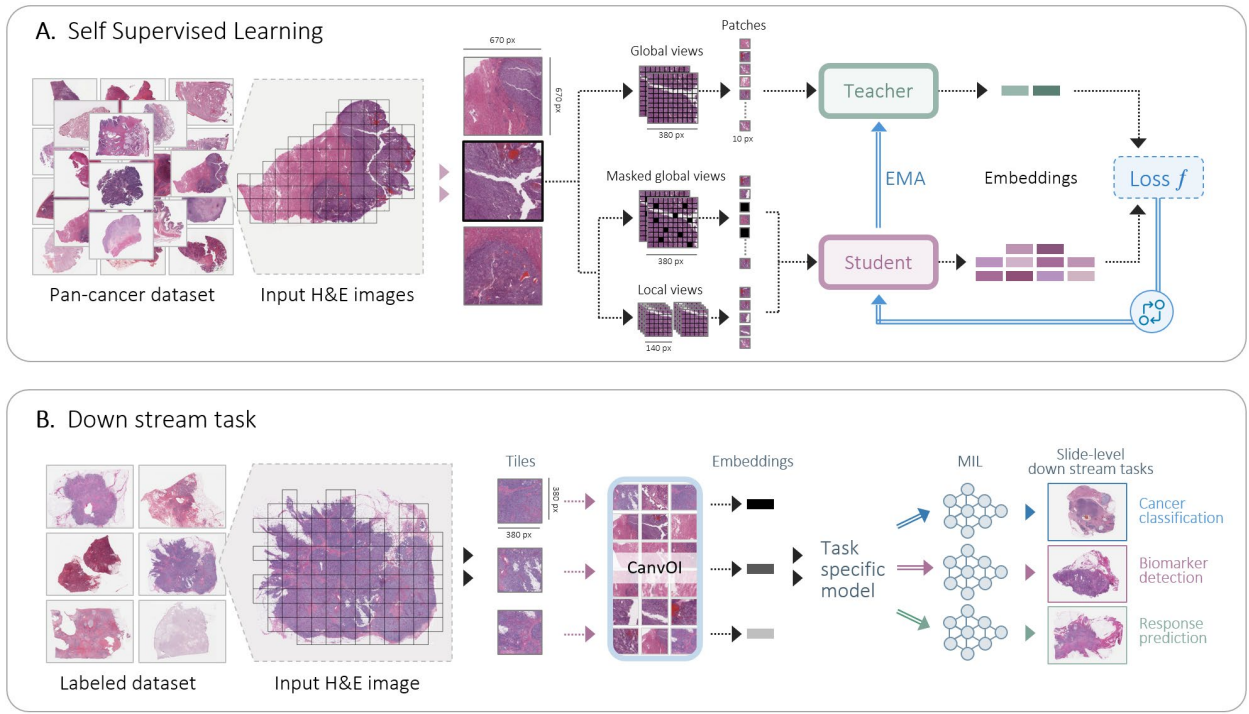

**Figure 2. Overview of CanvOI, a ViT-g/10 model with the DINOv2 framework.**
**A.** Training of CanvOI involves the extraction of both global and local views from WSIs of the pre-training pan-cancer dataset. Each of the views is divided into a sequence of $10^2$ pixel patches which is then encoded by a ViT-g architecture into embeddings.

Training was performed using the teacher-student distillation scheme, where the student network is trained to match the representations produced by the teacher network, which are stabilized by updating the teacher with an Exponential Moving Average (EMA) of the student's weights[9,10,23]. **B.** A schematic overview of the application of CanvOI in developing slide-level classifiers for digital pathology, utilizing Multiple Instance Learning (MIL), a weakly supervised method.

## Experimental Design

We benchmarked CanvOI as well as other publicly available pre-trained foundation models specifically designed for digital pathology, including H-optimus-0[31], Prov-GigaPath[14], Virchow[13] (we were unable to receive access to Virchow2) and Hibou-L[32]. The comparative Features of the foundation models used in this study are presented in figure 3.

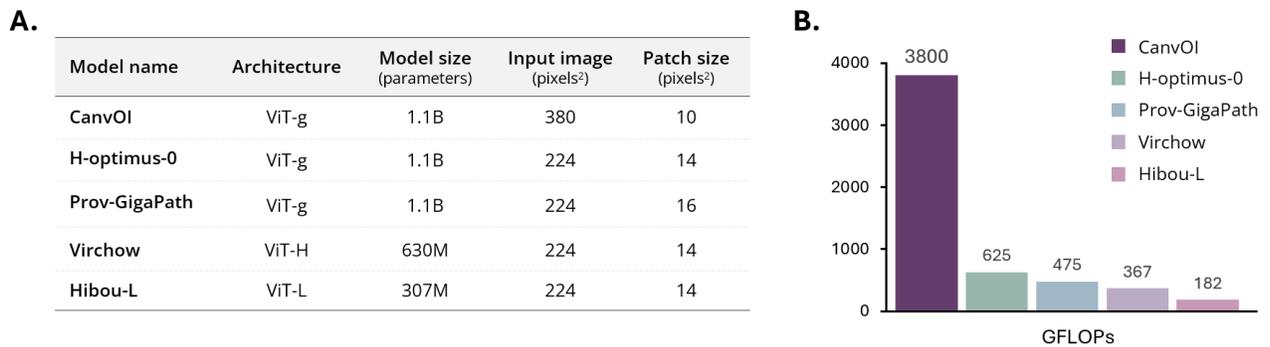

**Figure 3. Summary of model characteristics.**
**A.** Vision transformer specifications of foundation models used in this study.
**B.** Measured compute for one tile in Giga floating point operations (GFLOPs).

Weakly supervised whole slide image (WSI) classifications were performed using an attention-based multiple-instance learning (AB-MIL) algorithm[33]. For all tasks, 380 × 380 pixel tiles were used for CanvOI and 224 × 224 pixel tiles for all other models. All AB-MIL models were trained with a learning rate of $1\times10^{-4}$, weight decay of $1\times10^{-5}$, Adam optimizer[34] for 15 epochs. For the BRACS and HunCRC datasets, training was conducted using predefined training-validation-testing sets as described below. For the internal benchmark, which included images obtained from 3 source sites, a Leave-One-Group-Out cross-validation was used. In each iteration, 5-fold cross-validation on two sites, with label-stratification, created an ensemble model to test the third, out-of-domain (OOD) site. Area Under the Receiver Operating Characteristic Curve (AUC ROC, hereafter AUC) was used for performance evaluation. For tasks with multiple classes, macro-AUC ROC was employed.

### Slide-Level Downstream Tasks Using Public and Internal Datasets

**BRACS Dataset: Breast tissue histological classification**
The BReAst Carcinoma Subtyping (BRACS) dataset is composed of 547 H&E-stained formalin-Fixed Paraffin-Embedded (FFPE) breast tissue samples collected from 189 patients[35,36] (images and data available at https://www.bracs.icar.cnr.it/). WSIs were obtained from the National Cancer Institute IRCCS Fondazione G. Pascale, Naples, Italy and scanned using the Leica Aperio AT2 Digital Slide Scanner. The WSIs include benign, premalignant, and malignant samples. Two levels of classification tasks were performed. The first categorized samples into 3 lesion types: 'benign,' 'atypical,' or 'malignant.' The second subdivided these categories into 7 subtypes: 'benign' into 'normal', 'pathological benign' and 'usual ductal hyperplasia (UDH)'; 'atypical' into 'flat epithelial atypia (FEA)' and 'atypical ductal hyperplasia (ADH)'; and 'malignant' into 'ductal carcinoma in situ (DCIS)' and 'invasive carcinoma'. The dataset's official training, validation, and testing splits were used, comprising 395, 65 and 87 slides, respectively, for both the lesion type and lesion subtype tasks. Out of these, 2 slides were excluded, one from the training set as no tiles the size of $380^2$ could be extracted and one from the validation set as no tiles could be extracted.

**HunCRC Dataset: Colorectal tissue histological classification**
The Hungarian Colorectal Cancer Screening (HunCRC) dataset is composed of 200 H&E-stained FFPE colorectal tissue samples from Semmelweis University, Budapest[37] (images and data available at https://www.cancerimagingarchive.net/collection/hungarian-colorectal-screening/). The WSIs were obtained using a 3DHistech Pannoramic 1000 Digital Slide Scanner. The dataset includes normal, non-neoplastic, and neoplastic samples. The classification task categorized samples into 4 lesion types: 'normal', 'non-neoplastic lesions (NNL)', 'adenoma' and 'colorectal cancer (CRC)'. We followed the training, validation, and testing splits used by Chen et al.[12]. Samples were label-stratified into 50:25:25 folds, with training, validation, and testing sets comprising 100, 50 and 50 slides, respectively.

**Internal dataset: Lung cancer biopsy site and histological subtype classification**
The lung cancer dataset consists of 1,079 H&E-stained FFPE non-small cell lung carcinoma (NSCLC) WSIs from unique patients, obtained from either the primary or metastatic tumor sites, from our internal dataset. The WSIs were obtained from three source sites not included in the pretraining dataset, with 149, 489 and 441 WSIs from each site. Two Leica digital scanners were used: the Aperio AT2 and Aperio GT 450. Two binary classification tasks were performed. The first, using 1,079 WSIs, classified the biopsy site as either primary or metastatic. The second task classified the NSCLC histological subtype of 962 WSIs into lung adenocarcinoma (LUAD) or lung squamous cell carcinoma (LUSC), the two subtypes represented in this slide subset. We employed a Leave-One-Group-Out cross-validation scheme, where models developed during training were evaluated on samples from a source site excluded from the training data. In each of the three iterations, the dataset was divided into training and test sets based on the source site. During training we utilize a 5-fold cross-validation scheme to create ensemble of 5 models, normalized using z-score. This ensemble was then used for inference of the test set. For the evaluation, we merged the test sets from the three sites into a single cohort, consolidating the results from the various tests. We then calculated the AUC for the combined dataset across all sites.

**Impact of Labeled Data Reduction on Model Performances**

To evaluate the dependency on labeled data, we used the internal dataset consisting of 1,079 slides for the biopsy site task and 962 slides for the histological subtype task. We gradually reduced the number of slides used by the AB-MIL model in the cross-validation training step, with reductions to 50%, 30%, and 10% of the original training/validation sets. This reduction was performed using label-stratification, maintaining the percentage of each label during the training. Performance was then assessed using all slides in the Leave-One-Group-Out test sets. We evaluated the AUC averaged over all tasks (average AUC) at each reduction level.

# Results

CanvOI, a 1.1 billion parameters foundation model, was pre-trained on a dataset that included more than 70 million image tiles extracted from over 630 thousand tissue samples (Figure 1). We chose to train the ViT-g model within the DINOv2 self-supervised learning (SSL) framework using an image size of 380 x 380 pixels and a patch-size of 10, rather than the more common smaller tiles and patch-size of 14.

For the evaluation of CanvOI, we focused exclusively on slide-level prediction tasks. These tasks reflect the common clinical and diagnostic needs in digital oncopathology, where classifications of WSIs are typically required. For these tasks, we utilized AB-MIL, a weakly supervised method, to group the tile embeddings produced by the foundation model encoders into slide-level predictions. CanvOI embeddings were compared to those generated by other publicly available pre-trained foundation models specifically designed for digital pathology, including H-optimus-0[31], Prov-GigaPath[14], Virchow[13] and Hibou-L[32] across tasks involving breast, colorectal, and lung cancers.

The first two classification tasks involved breast cancer subtyping at two levels, one a higher level with 3 classes (benign, atypical and malignant) and a second, more detailed level with further subclassifications, using the BRACS benchmark and the predefined training/validation/test splits[35]. The next prediction task focused on colorectal cancer subtyping into four different categories (from normal to cancerous) using the HunCRC benchmark[37], with label-stratified training/validation/test cohorts. Lastly, using an internal benchmark, we classified non-small cell lung cancer (NSCLC) samples, a cancer type frequently used for benchmarking, through two binary classification tasks. The first task involved distinguishing between the major NSCLC histological subtypes, adenocarcinoma and squamous cell carcinoma. The second task classified the biopsy site, determining whether the tissue was taken from the primary tumor or a metastatic site. For these tasks, we used a Leave-One-Group-Out cross-validation with label-stratified training/validation splits (for more details see the Methods section).

Our model outperformed other pre-trained foundation models, demonstrating superior averaged AUC across all tasks, achieving 1.5-7.4% improvement compared with the other competing models (Figure 4). CanvOI excelled in at least 4 out of the 5 tasks when compared to individual models, highlighting its strong capability to extract meaningful information from pathological slides and outperform other foundation models across a range of evaluations.

One of the barriers that foundation models aim to overcome is the limited availability of labeled data. To assess the extent of dependency on labeled data, we utilized our internal lung benchmark, which includes approximately 1,000 slides. We evaluated the performance of different foundation models as the number of labeled slides gradually decreased, allowing us to observe how model effectiveness scales with varying amounts of labeled data. Figure 5 illustrates the AUC averaged over the NSCLC classification tasks. CanvOI achieved the highest performance with the largest margin when using 10% of labeled slides, maintaining top performance with the full set of slides, although with a narrower margin. Notably, CanvOI achieved an AUC of 0.83 on the smallest dataset, containing 10% of the dataset (only ~100 slides), exhibiting strong performance even with limited data, an essential feature for oncological research and clinical applications.

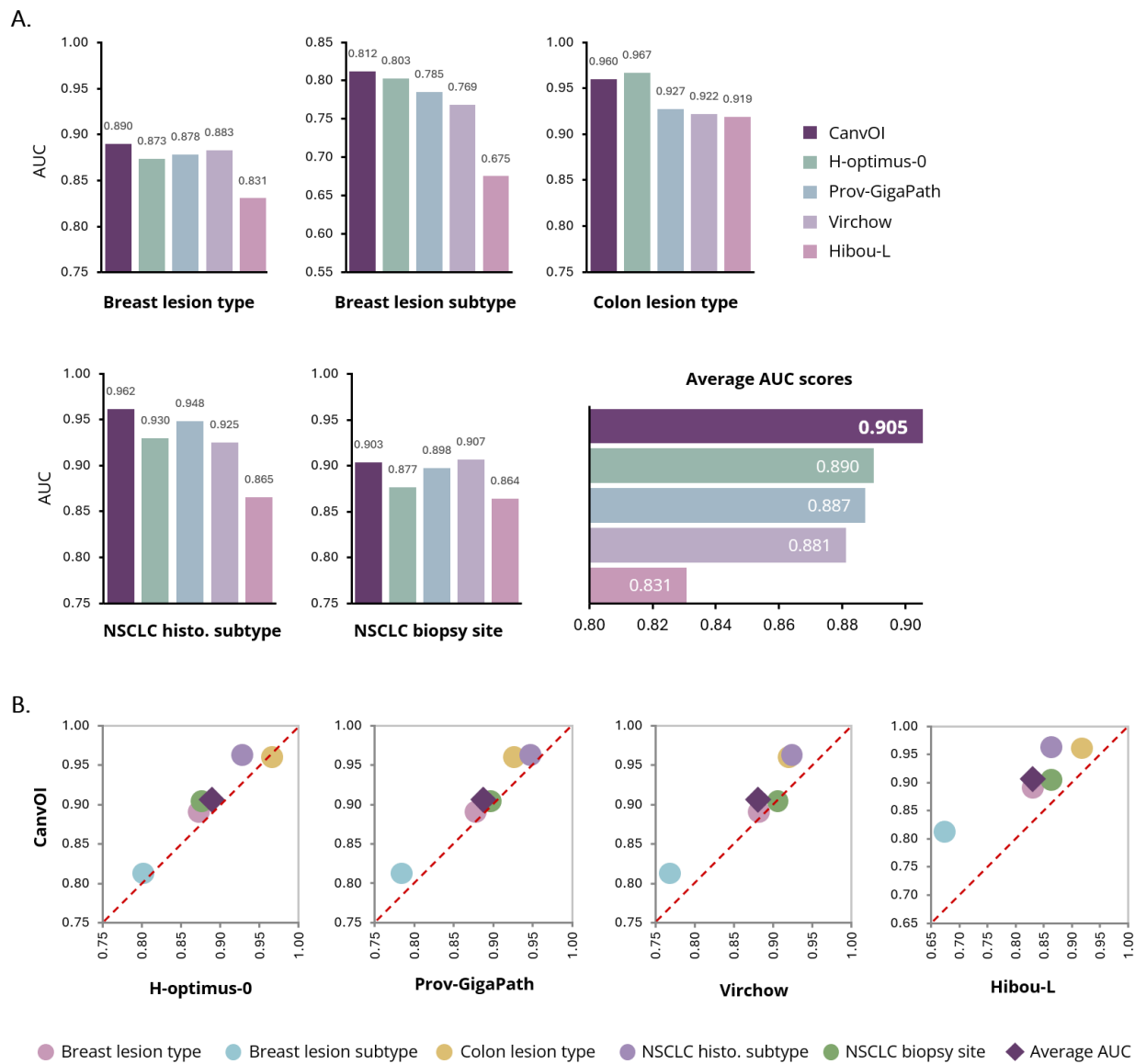

**Figure 4. Slide-level task performance.**
**A.** Performance of CanvOI and other comparator foundation models, H-optimus-O, Prov-GigaPath, Virchow and Hibou-L, across different downstream weakly supervised classification tasks and the average performance across all tasks. Presented are the AUC ROC of binary and macro-AUC ROC for multiclass classification tasks. **B.** A scatter plot comparing the performance between CanvOI and the comparator model in terms of AUC across all 5 evaluation tasks, as well as the average performance across all tasks.

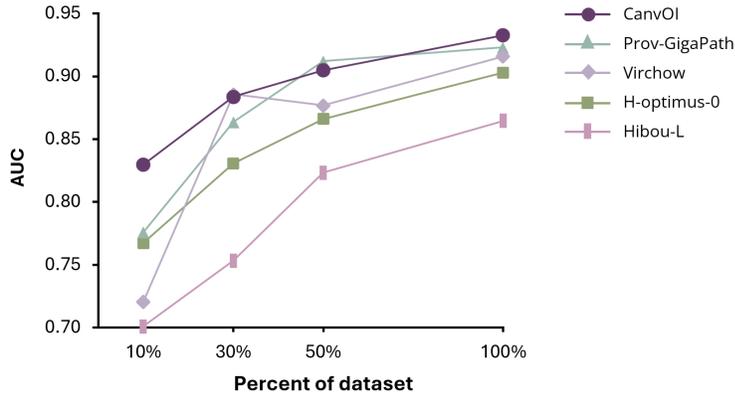

Figure 5. Performance across varying amounts of labeled data.
Averaged AUC of CanvOI and the comparator models over the NSCLC tasks (biopsy site and histological subtypes) as the number of labeled slides decreases. The 100% level represents approximately 1,000 slides.

## Discussion

Developing tools to support the rapidly evolving field of digital pathology is crucial for advancing research and enhancing diagnostic capabilities. A significant barrier is the high demand for addressing diverse questions in varied contexts coupled with the limited availability of labeled data. Large-scale foundation models are emerging as a potential solution to this challenge. However, optimizing computational resources in the realm of digital pathology foundation models is essential. With this in mind, we developed a state-of-the-art pathology image-based foundation model, also taking into consideration the nature of histological images and the limitations of MIL models. We hypothesized that increasing the tile sizes while decreasing the patch size, would yield embeddings that capture highly detailed information on the input image while reducing the number of embeddings fed to the downstream MIL models. This approach supports the creation of embeddings that reliably encapsulate the intricate details of histopathological images, ensuring a comprehensive and accurate representation suitable for downstream tasks without the need for increasing model size, which can be challenging to train[38].

We present CanvOI, a ViT-g/10 model developed using the DINOv2 self-supervised learning approach and an extensive and diverse pre-training dataset containing more than 70 million image tiles extracted from over 630 thousand tissue samples across more than 40 major organs and tissue types. In this study, we explored using 380 x 380 pixel tiles with a patch size of 10, unlike the traditional tile sizes of 224 × 224 pixels and patch size of 14/16, commonly used in natural image and current digital pathology models[12-14,16]. We believe that this approach is particularly relevant in digital pathology, where MIL models are often used for downstream classification tasks. In MIL, the model needs to process numerous tiles from each WSI and combine them into a single prediction, which can present challenges. There are ongoing efforts to develop improved aggregation models in the pathology domain[39-41] to tackle this challenge. A less explored area is the potential impact of altering the number of embeddings while maintaining the represented tissue constant. It would be valuable to extend the current study to further explore how such changes could influence the performance of aggregation models. Moreover, investigating the optimal compute utilization by balancing tile and patch sizes with model parameters to achieve stability during training and produce embeddings that support highly accurate downstream models could represent a new approach that has the potential to accelerate advancements in the field.

The evaluation of CanvOI on slide-level prediction tasks across various cancers, breast, colorectal and lung, demonstrated its robust performance. We chose to focus on slide-level tasks, which reflect real-world clinical needs, where the questions are at the slide or even patient level, such as detection of genomic alterations, assessment of prognosis and prediction of response to treatment and survival. We demonstrate that CanvOI excels in extracting meaningful information from WSIs, outperforming other pre-trained foundation models in most tasks. Overall, CanvOI achieved superior averaged AUC, showing a 1.5-7.4% improvement compared to the other foundation models (Figure 4). Moreover, our results demonstrate that CanvOI significantly outperformed the other models, with the performance gap widening substantially when trained on just 10% of the initial cohort (approximately 100 slides) (Figure 5). This demonstrates that our approach is well-suited to the limitations commonly encountered in clinical settings, where access to large amounts of labeled data may be restricted. This could be particularly valuable in clinical trial settings, where limited numbers of participants often inhibit the ability to extract meaningful outcomes from the data. Foundation models like CanvOI can support analysis and interpretation of trial results, enhancing the insights derived from the data and contributing to more successful outcomes.

In conclusion, CanvOI marks a significant advancement in the development of foundation models for histopathology. By integrating extensive data diversity and substantial model size coupled with the introduction of new tile and patch sizes, our approach sets a new benchmark in the field. We believe that further refinement of this methodology will support the rapidly evolving field of foundation models, which will undoubtedly play a pivotal role in the future of digital pathology, laying the building blocks for Oncology Intelligence (OI), enhancing clinical outcomes of patients with cancer, accelerating medical research and fostering new discoveries.

## Acknowledgements

We would like to thank the teams at Imagene for their invaluable support and work, without which this work would not have been possible. A special thanks goes to Avital Rabani for graphics and editing support. The results published here are partly based upon data generated by The Cancer Genome Atlas (TCGA) Research Network, the Clinical Proteomic Tumor Analysis Consortium (CPTAC), the National Cancer Institute's Cancer Moonshot Biobank (CMB) and the Genotype–Tissue Expression (GTEx) consortium.

**CanvOI: Scaling FLOPS differently**